\documentclass[12pt, draftclsnofoot, onecolumn]{IEEEtran}
 \usepackage[dvips]{graphicx,psfrag}
\usepackage{amsthm}
\usepackage{amsmath}
\usepackage{amsmath}
\usepackage{amssymb}
\usepackage{color}
\ifCLASSINFOpdf
 
\else
 
\fi

\begin{document}

 \title{A Note on a Sum of Lognormals}

\author{By N B Chakrabarti,\\ ECE Dept., IIT Kharagpur, Kharagpur, India. 
        
}


\maketitle

\begin{abstract}
This note considers the applicability of Gauss-Hermite quadrature and direct numerical quadrature for computation of moment generating function (mgf) and the derivatives. A preprocessing using the asymptotic technique is employed while computing the characteristic function (chf) using Gauss – Hermite quadrature while this is optional for mgf. The mgf of the low and high amplitude regions of a single lognormal variable and the derivatives is examined and attention is drawn to the effect of variance. The problem of inversion of the mgf/chf of a sum of lognormals to obtain the CDF/pdf is considered with special reference to methods related to Post Widder technique, Gaussian quadrature and the Fourier series method. The method based on  the complex exponential integral which makes use of the derivative of the cumulant or non-harmonic Fourier series is an alternative. Segmentation of the mgf/chf on the basis of the derivative structure which indicates activity rate is shown to be useful.
\end{abstract}


\IEEEpeerreviewmaketitle


\section{Introduction}
\IEEEPARstart{T}{he} behavior of a sum of lognormals has been extensively studied [1-10]. The Fenton-Williamson (FW) technique of representing the distribution of the sum in terms of an equivalent lognormal remains a powerful tool. It is known that moderation of moment values by truncation enhances the range of applicability of the FW method. An alternative of finding the mgf/chf of the independent lognormals together with other relevant distributions to compute the product mgf/chf and then finding the inverse has also been studied [8,9]. The present work is concerned with this approach.

A random positive variable $x$ is log-normally distributed if the logarithm of $x$ is normally distributed. Its probability density function (pdf) and cumulative distribution function (CDF) are:
\begin{equation}
 p(x)=\frac{1}{x\sigma\sqrt{2\pi}}\exp\left(-\frac{\left(\ln{x} -\mu\right)^2}{2\sigma^2}\right)
 \label{Eq:lognormal_pdf}
\end{equation}
and 
\begin{equation}
 F(x)=\frac{1}{2}+\frac{1}{2}\text{erf}\left[\frac{\ln{x}-\mu}{\sqrt{2\sigma}}\right]
 \label{Eq:lognormal_CDF}
\end{equation}
It is easy to verify that $\ln(p(x))$ is quadratic in $\log(x)$; this is used to test lognormality. The fact that $p(x)dx$ equals $-p(y)dy$ where  $y=\frac{1}{x}$ is the basis of reduced range integration. In the text, the parameter $\mu$ is suppressed and is restored in the final step. 
\section{Computation of Moment Generating Function}
The moment generating function $M(s)$ of a lognormal random variable and its derivatives can be found using Gauss Hermite quadrature. For even length 
\begin{equation}
 M(s)=\sum\limits_{k=-N}^Nw_k\exp\left(-s\cdot\exp\left(\sqrt{2}\sigma x_k\right)\right)
 \label{Eq:lognormal_mgf_GH_quadrature}
\end{equation}
where $x_k$ are the abscissas and  $w_k$ are the weights. The transform variable $s$ will be assumed to be real except when otherwise specified. 

One may express $M(s)$ as a sum of contributions $M_1(s)$ and $M_2(s)$ from the regions where $0\leq x\leq1$ and $x\geq 1$ respectively, i.e., 
\begin{align}
 M_1(s)&=\sum\limits_{k=1}^Nw_k\exp\left(-s\cdot\exp\left(-\sqrt{2}\sigma x_k\right)\right)\\
 \text{and~~~}M_2(s)&=\sum\limits_{k=1}^Nw_k\exp\left(-s\cdot\exp\left(\sqrt{2}\sigma x_k\right)\right)
\end{align}

These are equal at $s=0$ and comparable in magnitude at very low frequencies. Both $M_1$  and $M_2$ decrease with $s$, $M_2$ faster than $M_1$. 
 The $n$-th derivative of $M(s)$ is
 \begin{equation}
  M_n(s)=(-1)^n\sum\limits_{k=-N}^Nw_k\exp\left(n\sqrt{2}x_k\right)\exp\left(-s\cdot\exp\left(\sqrt{2}\sigma x_k\right)\right)
  \label{Eq:nth_derivative_mgf_GH}
 \end{equation}
Equation \eqref{Eq:nth_derivative_mgf_GH} shows that the equivalent weights $w_k\exp(n\sqrt{2}x_k)$ for derivatives are smaller for $M_1(s)$. The contributions from negative $x_k$, i.e.,  $x$ lying within the region from zero to unity, is therefore initially small but the decay with $s$ is slow. 

Holgate [6] used an asymptotic technique to derive the closed form expression for chf  $\varphi(j\omega)$ as
\begin{equation}
 \varphi(j\omega)=\frac{1}{\sqrt{1-z}}\exp\left(-\frac{z^2-2z}{2\sigma^2}\right)
 \label{Eq:chf_Holgate}
\end{equation}
The parameter $z$ satisfies the Lambert  W equation  $z\exp(-z)=j\omega\sigma^2$ [9,10].

Equation \eqref{Eq:chf_Holgate} is known to be applicable at high frequency and small value of variance. The saddle point method enables one to compress the frequency scale by a transformation so that $z$ is restricted to lie in the second quadrant. The applicability of this method can be improved if one expresses the chf as a product of $H(z)=\exp\left(-\frac{z^2-2z}{2\sigma^2}\right)$ and a term  $G(z)$ derived by integrating the residual employing Gauss-Hermite quadrature.

An expression for $M(s)$ based on the asymptotic technique and Gauss Hermite quadrature may be written as 
\begin{align}
 M(s)=\exp\left(-\frac{\left(W^2+2W\right)}{2\sigma^2}\right)\sum\limits_{k}w_k\exp\left(-\frac{W}{\sigma^2}\left(\exp\left(\sqrt{2}\sigma x_k\right)-\sqrt{2}\sigma x_k-1\right)\right)
 \label{Eq:Asymptotic_GH_mgf}
\end{align}
where $W$ satisfies the  Lambertw equation, viz., $W\exp(W)=s\sigma^2$. An equivalent form is 
\begin{equation}
 M(s)=\exp\left(-\frac{W^2}{2\sigma^2}\right)\sum\limits_k w_k \exp\left(-\frac{Wg(x,\sigma)}{\sigma^2}\right)
 \label{Eq:Asymptotic_GH_mgf_a}
\end{equation}
where $g(x,\sigma)=\exp\left(\sqrt{2}\sigma x\right)-\sqrt{2}\sigma x$.

An examination of Equations \eqref{Eq:lognormal_mgf_GH_quadrature} and \eqref{Eq:Asymptotic_GH_mgf} is instructive. For practical applications the number of terms necessary does not usually exceed twenty. When the frequency is high, one would require still fewer terms. 
$W$ is almost linear with $s$ at small $s$.  One can replace the exponent in the RHS of \eqref{Eq:Asymptotic_GH_mgf_a} by $\left(1-\left(1-\exp(-s)\right)\right)^{g(x,\sigma)}$. $M(s)$ can then be written as $\exp\left(-W^2\right)P_1(s)$, where $P_1(s)$ is a polynomial in $s$. For high frequencies on the other hand, $W$ varies logarithmically and one obtains an approximation $M_h(s)\exp\left(-\frac{\ln^2(s)}{2\sigma^2}\right)P_h\left(\frac{1}{s}\right)$, where  $P_h\left(\frac{1}{s}\right)$ is a polynomial in $\frac{1}{s}$. These point to the applicability of Pade approximation.

MGF and its derivatives can also be computed using simple quadrature in the reduced range of $x$ from zero to unity. These can be written as:
\begin{align}
 &M(s)=\int\limits_0^1 p(x)\left(\exp\left(-sx\right)+\exp\left(-\frac{s}{x}\right)\right)dx \label{Eq:mgf_and_derivative_1}\\
 &\frac{d^kM(s)}{ds^k}=(-1)^k\int\limits_0^1 p(x)\left(x^k\exp\left(-sx\right)+\left(\frac{1}{x}\right)^k\exp\left(-\frac{s}{x}\right)\right)dx
 \label{Eq:mgf_and_derivative_2}
\end{align}
It is evident from Equation \eqref{Eq:mgf_and_derivative_1} that for zero mean lognormal, inverse moments are equal to the positive moments.

\begin{figure}[t]
\begin{center}
\includegraphics[width=.6\textwidth]{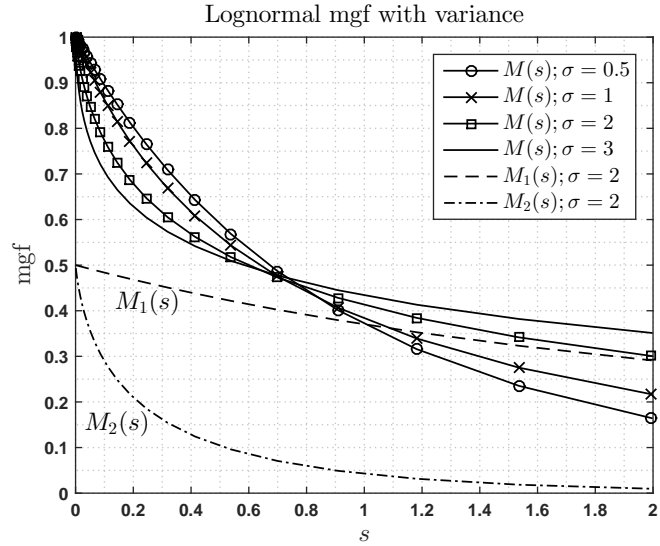}\vspace{-.3cm}
 \caption{Moment generating function for a set of variances. Relative contributions from $M_1$ and $M_2$ are also indicated.} \vspace{-.3cm}
\label{Figure1}
\end{center}
\end{figure}
If the variance is large, $M(s)$ decreases fast at low frequencies and has a knee like appearance. This results in the crossing of the mgfs of different variances typically close to $s=0.6$ as shown in Figure \ref{Figure1}.
This signifies that moment or cumulant matching techniques must avoid regions close to where mgf corresponding to a sum of lognormals may not possess uniqueness. One notices that the mgf of a sum of lognormals has a wide variety of descent patterns depending on the distribution of the variances. The variety becomes wider when Rice/Suzuki and other distributions are also associated.

\begin{figure}[t]
\begin{center}
 \includegraphics[width=.6\textwidth]{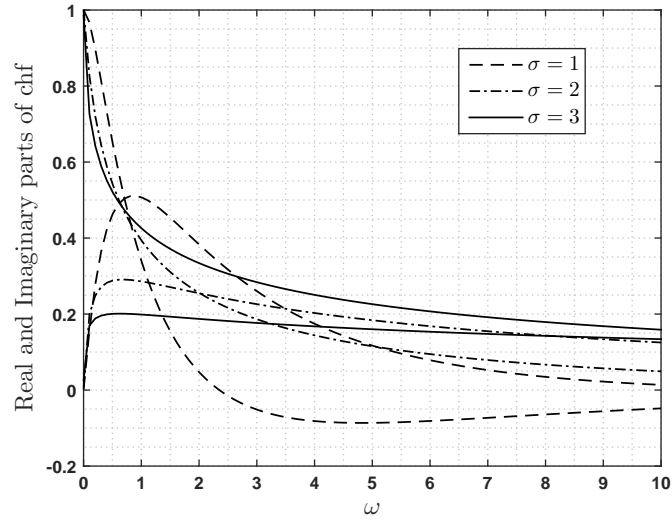}\vspace{-.3cm}
 \caption{Real and imaginary parts of chf for a set of variances.} 
\label{Figure2}
\end{center}
\end{figure}

Figure \ref{Figure2} shows the variation of real and imaginary parts of the chf for different variances. Here again crossings occur for the real and imaginary parts but at different locations.   The spread is wider compared to that of mgf.

\begin{figure}
\begin{center}
 \includegraphics[width=.6\textwidth]{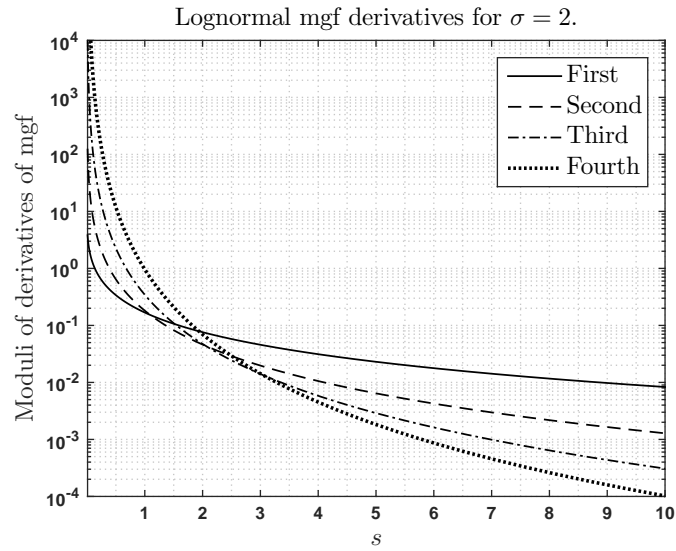}
 \caption{Variation of first four derivatives w.r.t. frequency ($s$) for lognormal mgf.} 
\label{Figure3}
\end{center}
\end{figure}

The derivatives of the mgf convey a wealth of useful information. The values of the derivatives are known to be large at $s=0$; they give the moments. These decrease fast with s becoming negligible beyond $s=2$. Figure \ref{Figure3} shows the variation of the first four derivatives with $s$. It is often assumed that the estimate of equivalent lognormal may be improved by using derivatives higher than two but one must note that these lose their dominance as the frequency increases. In fact if one considers only the term $\exp\left(-\frac{\left(W^2+2W\right)}{2\sigma^2}\right)$, the inverse of the equivalent variance equals the sum of the inverses of individual variances. 

 An analytical expression for mgf at very high frequencies has been obtained by Barouch and Kaufman [5]. Letting $z=sx$, the integral for $M(s)$ becomes
 \begin{equation}
  M(s)=\frac{1}{\sqrt{2\pi}\sigma}\int\nolimits_0^\infty \frac{1}{z}z^{\frac{\ln s}{\sigma^2}}\exp\left(-z-\frac{\ln^2s}{2\sigma^2}-\frac{\ln^2z}{2\sigma^2}\right)dz
  \label{Eq:Integral_mgf}
 \end{equation}
Ignoring correction terms for simplicity, the integral may be approximated by  
\begin{equation}
 M(s)=\frac{1}{\sqrt{2\pi}\sigma}\exp\left(-\frac{\ln^2s}{2\sigma^2}\right)\Gamma\left(\frac{\ln s}{\sigma^2}\right)
\end{equation}
The derivative of order $n$ may be obtained approximately by augmenting the argument of the Gamma function by $n$ and dividing the result by $(-s)^n$. 

A study of the derivative enables one to understand why some techniques for estimating the behavior of lognormal work best in some regions. A knowledge of derivatives is useful in controlling the sampling rate in the regions where $M(s)$ varies fast. An immediate application is to improve the accuracy of inversion by incorporating interpolation in regions where derivatives are large. For a sum of  lognormals, the first derivative is not hard to calculate; one merely multiplies the mgf by the sum of logarithmic derivatives of individual lognormals. Numerical differentiation is also an alternative. The fact that the derivatives at low frequencies are dominated by large $x$ region of pdf while those at high frequencies are contributed by the small $x$ region indicates that components on two sides of the median can be separated by high order differentiation or differencing as in Post-Widder formula . 
The behavior of a lognormal sum depends as is well known on the distribution of variances. When the combination of large variance components has amplitude lower than that of the combination of low variance components, the sum has a uniformly decaying character. In a situation where  the mgf of low variance components is comparable to or lower than the mgf of the large variance components in a significant part of the spectrum, a variety of patterns emerges. 
\section{Inversion Methods for Deriving CDF, pdf and derivative}
Once the mgf/chf of the individual components has been found, an important task is the computation of the CDF and pdf corresponding to the mgf/chf of the product.

In many applications, a knowledge of CDF is the primary objective. This is found by integrating the contributions from the entire frequency domain. One is often interested in the information about lower tail, the upper tail and the region close to the median. In such cases the computational load can be considerably reduced.  Techniques of inversion which yield both CDF and pdf find  application in the  estimation of median starting from the value obtained from FW's technique.

Laplace transforms Inversion Methods: A large number of methods for numerically inverting Laplace transforms have been developed [12, 13]. These may be classified according as they employ real and complex arithmetic derived respectively from the Post Widder technique and Bromwich contour integral. The survey of Brian and Martin [12] groups them as: (a) methods which compute a sample, (b) methods which expand in exponential functions, (c) Gaussian numerical quadrature, (d) representation as Fourier series based on the Poisson summation formula and (e) Pade approximation. The Talbot technique based on deforming the contour deserves mention. Widder  in his book [11] states an asymptotic formula for the inverse 
\begin{equation}
 f(x)=\left(-1\right)^k\left(\frac{k}{x}\right)^{k+1}\frac{1}{k!}M^k\left(\frac{k}{x}\right)
\end{equation}
where $M^k$ is the $k$-th derivative of the input transform $M(s)$ at $s=\frac{k}{x}$. This formula is known to converge slowly. Der Haar [12] has shown that in some cases the simple formula $\frac{1}{x}M\left(\frac{1}{x}\right)$  gives reasonable answers. This and the variant $\frac{1}{2x}M\left(\frac{1}{2x}\right)$ are single point formulas.Better approximations are provided by the formulas using derivatives; while these are not accurate, they serve as indicators of expected comparisons.In many cases the expression based on the first derivative proves useful. Post-Widder formula is remarkable in establishing direct relation between $M(s)$ and $f(x)$. Davies and Martin [12] point out a relation between Post Widder and a delta-convergent sequence.  
\subsection{Gaver and Gaussian Quadrature:} Gaver [14] developed a family of three parameter functions
\begin{equation}
 f(n,m,a)=\frac{(n+m)!}{n!(m-1)!}\left(1-\exp(a\mu)\right)^n\exp\left(ma\mu\right)
\end{equation}
where $m$ and $n$ are integers and the parameter $a$ is inversely proportional to $x$. Gaver-Stehfest inversion technique originated from this formula based on the difference operator in the special case where $m=n$. This may be written as
\begin{equation}
 f(x)=\frac{a}{x}\sum K_nM\left(\frac{na}{x}\right)
 \label{GaverStehfest}
\end{equation}
where $a=\ln 2$ and $K_n$  are the coefficients and the summation is carried out over $N$ terms. A similar relation was derived by Zakian [16] where $a=2$. It was shown that the condition required for finding CDF is satisfied. Gaussian numerical quadrature of the inversion integral has yielded a large class of methods using orthogonal polynomials which give approximate formula for inversion which is exact when $M(s)$ is a linear combination of inverse powers of s up to an order $2(N-1)$ (Piessens, [15]). Many variants of the Gaussian quadrature technique have been developed [12,13], and nodes and weights for computing the inverse tabulated. The inversion integral is first written as
\begin{equation}
 f(x)=\frac{1}{2\pi jx}\int M\left(\frac{z}{x}\right)\exp(z)dz
\end{equation}
The integral is approximated by a sum.
\begin{equation}
 f(x)=\sum \frac{1}{x}K_n M\left(\frac{a_n}{x}\right)
 \label{equation18}
\end{equation}
and the coefficients are computed. The roots $a_n$ are complex and so are the coefficients.

A simple way to obtain $a_n$ is to replace the exponential function (i.e. $\exp(z)$) by the Pade approximation [12] and find the roots of the denominator polynomial. CDF is obtainable as
\begin{equation}
 F(x)=\sum \frac{K_n}{a_n} M\left(\frac{a_n}{x}\right)
\end{equation}
The $k$-th derivative is given by 
\begin{equation}
 P_k(x)=\sum\frac{1}{x}K_n\left(\frac{a_n}{x}\right)^kM\left(\frac{a_n}{x}\right)
\end{equation}
\subsection{Fourier Series Method:}
The Fourier series method is popular in statistical literature. 
The simpler Fourier inversion form uses either the cosine transform or the sine transform, where the cosine transform is given by
\begin{equation}
 f_{\text{c}}(x)=\frac{2}{L}\exp\left(\frac{cx}{L}\right)\sum\limits_k\Re\left(\frac{c+j\pi k}{L}\right)\cos\left(\frac{\pi k x}{L}\right)
\end{equation}
and  the sine transform is given by
\begin{equation}
 f_{\text{s}}(x)=-\frac{2}{L}\exp\left(\frac{cx}{L}\right)\sum\limits_k\Im\left(\frac{c+j\pi k}{L}\right)\sin\left(\frac{\pi k x}{L}\right)
\end{equation}
respectively, where $c$ defines the contour and $L$ is the length.  An equally weighted sum is known to achieve better accuracy.

\begin{figure}
\begin{center}
 \includegraphics[width=.6\textwidth]{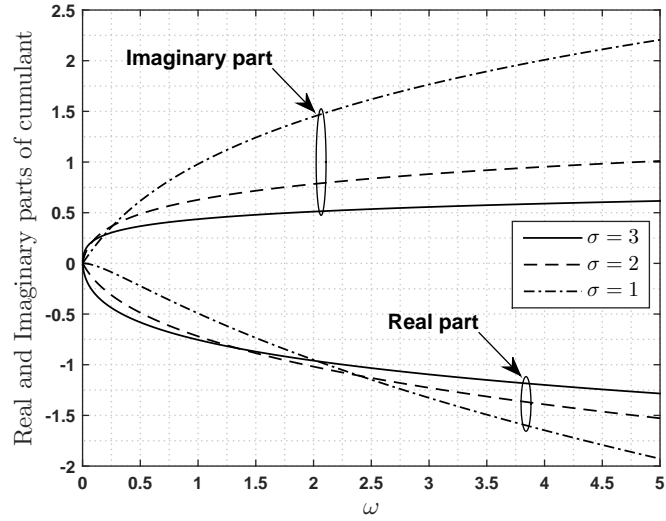}
 \caption{Cumulants of a chf of lognormal for a set of variances.} 
\label{Figure4}
\end{center}
\end{figure}

The corresponding expression for CDF due to Gil-Pelaez [17,18,19] in terms of the chf $\varphi(j\omega)$ is 
\begin{equation}
 F(x)=\frac{1}{2}-\frac{1}{\pi}\int\limits_0^\infty \left[\exp\left(-j\omega x\right)\varphi(j\omega)-\exp\left(j\omega x\right)\varphi(-j\omega)\right]\frac{d\omega}{\omega}
\end{equation}
When the cumulants $\mathcal{X}_1$ and $\mathcal{X}_2$ are known this may be written as  
\begin{equation}
 F(x)=\frac{1}{2}-\frac{1}{\pi}\int\limits_0^\infty\exp\left(\mathcal{X}_1\right)\frac{\sin\left(\mathcal{X}_2-\omega x\right)}{\omega}d\omega
 \label{eq:Fx}
\end{equation}
Figure \ref{Figure4} shows the variation of cumulants with frequency ($\omega$) for three values of the variances.
The envelopes are found to cross near $\omega=2$.
Davies [18] presented a numerical technique for implementing the Gil-Pelaez formula for evaluating CDF as
\begin{equation}
 F(x)=\frac{1}{2}-\frac{1}{2}\sum\limits_{k=0}^N\Im\left\{\varphi\left(k+\frac{d}{2}\right)\frac{\exp\left(-j\left(k+\frac{d}{2}\right)x\right)}{\pi\left(k+\frac{d}{2}\right)}\right\}
\end{equation}
where $d$ is the grid spacing. 
\subsection{Complex exponential integral:}
It is noticed from Figure \ref{Figure4} that the cumulant has a smooth characteristic except near the origin. A knowledge of the first derivatives of $\mathcal{X}_1, \mathcal{X}_2$ enables the use of the complex exponential integral [21,22] to compute the contributions to the CDF from different sub-intervals. For the purpose it is necessary to examine the nature of variation of the real and imaginary parts of the cumulant with frequency. The real part ($\mathcal{X}_1$) varies very slowly at low frequencies and the derivative is consequently very small. The derivative has a peak at about $\omega=0.01$ for a lognormal of variance $2$ and decreases monotonically thereafter. The derivative of the imaginary part has a large value at low frequencies and also decreases monotonically. The derivative of the real part is smaller in magnitude than that of the imaginary part at low frequencies and the role is reversed at high frequencies.

The complex exponential integral [21,22] is written as
\begin{equation}
\label{eq:EponentialInt}
 \mathbf{E}_1(z)= \int_z^\infty \frac{\exp(-t)}{t}dt
\end{equation}
Three relevant integrals are [21,22]
\begin{align}
 \int_0^1\frac{e^{-at}\sin bt}{t}dt&=\tan^{-1}\frac{b}{a}+\Im\mathbb{E}_1(a+jb)\label{eq:ComplexInt_1}\\
 \int_0^1\frac{e^{at}\sin bt}{t}dt&=-\tan^{-1}\frac{b}{a}+\Im\mathbb{E}_1(a+jb)+\pi \label{eq:ComplexInt_2}\\
 \int_0^1\frac{e^{-at}(1-\cos bt)}{t}dt&=\frac{1}{2}\log\left(1+\frac{b^2}{a^2}\right)-\Re\mathbb{E}_1(a)+\Re\mathbb{E}_1(a+jb)\label{eq:ComplexInt_3}
\end{align}
When the cumulant has linear variation with an attenuation constant \textquoteleft $a$\textquoteright~ and group delay \textquoteleft $b$\textquoteright~ in the interval of interest, the exponent of the integral $\varphi\left(j\omega\right)\exp(-j\omega x)/\omega$ may be written as $\mathcal{X}_1+j\mathcal{X}_2 - a\omega+j (b-x)\omega$ where $\mathcal{X}_1$ and $\mathcal{X}_2$ are the components of the cumulant at the lower end. In the interval of $\omega$ from zero to unity, the integral is given by [21,22] 
 \begin{equation}
\text{J}_\text{1}=\arctan\left(\frac{b-x}{a}\right)+\Im\text{E}_\text{1}\left(a+j(b-x)\right) 
\label{eq:J1}
\end{equation}
as $\mathcal{X}$ is 0 at the origin. At the high frequency end, the amplitude is small in the stretch extending from $\omega_m$  to infinity, the integral is
\begin{equation}
 \text{J}_\text{m}=\exp(\mathcal{X}_m)\text{E}_\text{1}(\omega_m(a_m+j(b_m-x)))\
 \label{eq:Jm}
\end{equation}

In the piecewise sum approach one sums the contributions from the different segments.To derive the value of the integral from $\omega_1$  to $\omega_2$, one subtracts the contribution from the ranges $\omega_1$  to infinity and $\omega_2$ to infinity with the appropriate values of the ($a$, $b$) parameters. One can include the  effect of the second derivative partially by expressing the difference as a short length Fourier series, i.e,
\begin{equation}
 \exp\left(\mathcal{X}_1(\omega)+a\omega-jb\omega\right)=1+\sum c_k\sin\left(\frac{\pi k\omega}{L}\right)
\end{equation}
where $L$ is the length of the segment. The resulting expression can be written as a sum of complex exponentials.

A substitution of the first term of the R.H.S of equation \eqref{eq:J1} in equation \eqref{eq:Fx} gives
\begin{equation}
 F_0(x)=\frac{1}{2}-\frac{1}{\pi}\arctan\left(\frac{b-x}{a}\right)
\end{equation}
The value at $x=0$ is different from the desired value of zero. On the other hand if one uses the relation for CDF based on the real part one obtains the integral 
\begin{equation}
 \int \exp(-a\omega)\cos(b\omega)\frac{\sin(\omega x)}{\omega} dw\nonumber
\end{equation}
which gives
\begin{equation}
 F_0(x)=\frac{1}{\pi}\arctan\left(\frac{b+x}{a}\right)-\frac{1}{\pi}\arctan\left(\frac{b-x}{a}\right)
 \label{eq34}
\end{equation}
The result is independent of the polarity of $b$.
Combining the $\arctan$ terms one obtains
\begin{equation}
 F_0(x)=\frac{1}{\pi}\arctan\left(\frac{2ax}{a^2+b^2-x^2}\right) 
\end{equation}
This shows that at low amplitudes $F_0(x)$ is linear. The median corresponding to $F_0(x)=0.5$ is given by $\sqrt{a^2+b^2}$. The pdf corresponding to \eqref{eq34} is $\frac{1}{\pi}\left[\frac{a}{a^2+(b+x)^2}+\frac{a}{a^2+(b-x)^2}\right]$. This has finite value at $x$ close to zero. This indicates that weighting at small values of $a$ and $b$ is necessary. This conclusion also follows from an examination of RHS of equation \eqref{eq:Fx}.

As an alternative to the piecewise sum which requires a precise knowledge of the local derivatives, one may express the chf as a sum of complex exponentials
\begin{equation}
 \varphi(j\omega)=\sum A_k\exp(-a_k\omega+jb_k\omega)
 \label{eq:damping}
\end{equation}

where $A_k$ is the amplitude of the k-th component. The CDF is then expressed as
\begin{equation}
 F(x)=\frac{1}{\pi}\sum A_k\left(\arctan\left(\frac{b_k+x}{a_k}\right)-\arctan\left(\frac{b_k-x}{a_k}\right)\right)
 \label{Eq:CDFExpInt}
\end{equation}

As mentioned earlier the absolute and relative values of  $a_k$ and $b_k$ depend on the frequency regions concerned. Specifically small $a_k$ occurs at very low frequencies for a short while and both $a_k$ and $b_k$ are small at very low frequencies. The CDF $F_0(x)$ rises fast if $a_k$ is small and slowly for large $a_k$ and $b_k$ is a shift parameter. There is a one to one correspondence between the frequency and the $a$, $b$ parameters.

A remark on exponential sum representation is relevant. There are methods of Laplace transform inversion which assumes the result can be expanded as a sum of exponential functions [12,13]. The Carath\`eodory representation has a form
\begin{equation}
 c_k=\sum_{m}\rho_m\exp(j\pi\theta_mk),
 \label{eq:Caratheodory_represetation}
\end{equation}
where $-1<\theta_m\leq 1$ and $\rho_j>0$. This yields on inversion a sum of step functions. Complex exponential sum is widely used in spectral analysis and other applications [23]. A common technique is to employ SVD on the frequency data and find roots of an eigen polynomial derived from it and then find the weights. This technique is not directly applicable because some roots give positive real part of the logarithm and the weights have complex values. These are not appropriate for the present application because the integrals that apply are then as in equation \eqref{eq:ComplexInt_2} and equation \eqref{eq:ComplexInt_3}.

In the present work, the  basis of selection of the density of nodes is the values of  the derivatives of the real and imaginary parts of the cumulant. As noted earlier the square root of $a^2+b^2$ yields the median of the individual arctangent graph. A choice of node interval proportional to the inverse of the local median seems appropriate. When $a$ and $b$ are both very small, the rise occurring at small $x$ is fast. The values of $x$ for $F_0(x)=0.1$ and $F_0(x)=0.9$ can be found from the arctan expression. The above remarks provide general guidelines but one must compare the values of the cumulant of chf computed from the expression for the exponential sum, insert/delete nodes and use weighting as necessary to correct the error using non-linear least square technique.

\begin{figure}
\begin{center}
 \includegraphics[width=.6\textwidth]{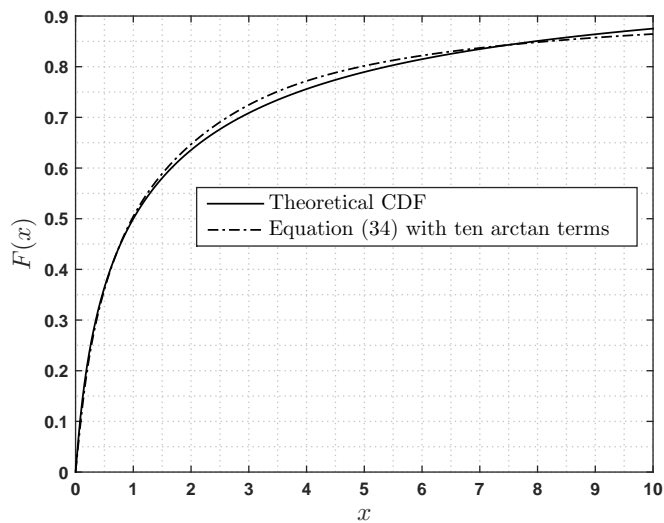}
 \caption{Comparison between theoretical CDF and computation using sum of $\arctan$s for $\sigma=2$.} 
\label{Figure6}
\end{center}
\end{figure}

Fairly accurate results are obtained even for an equally weighted sum if the contributions from very low, low, medium and high frequencies are included. It is noticed that each term in equation \eqref{Eq:CDFExpInt} has the appearance of a CDF increasing monotonically from zero to unity. Figure \ref{Figure6} compares a result with the theoretical CDF. It is necessary in applications of complex exponential integral to verify that the frequency behaviour of the sum is close to that of CHF. 

The first two derivatives  can be used for computing the pdf of a lognormal. If $a_2$ and $b_2$ are the second derivatives of the cumulants while $a_1$ and $b_1$ are the first derivative valid for a frequency segment, the chf may be written as
\begin{equation}
 \varphi(j\omega)=\sum A_k\exp\left(-(a_{2k}\omega^2+a_{1k}\omega)+j(b_{2k}\omega^2+b_{1k}\omega)\right)
\end{equation}
To compute the pdf, one may make use of the second derivatives of the cumulant together with the first derivatives. 

\textbf{The converse - chf from CDF:} If chf is expressed as a non-harmonic Fourier series (Paley and Wiener/Gil-Pelaez [17]/Carath\`eodory)  viz., 
\begin{equation}
 \varphi(j\omega)=\sum_{k=1}^NA_k\exp\left(j\omega x_k\right)
\end{equation}
the CDF derived from the real part is a staircase function with steps of $A_k$ at the nodes $x_k$. When one introduces a damping parameter to ensure the smoothness, the similarity with Eq. \eqref{eq:damping} becomes obvious. The converse of the above can be used to find the chf from the known values of the distribution function, for example weibull, lognormal and generalized gamma. When the random variables are uncorrelated one can find the chf of the sum by forming the product of individual chfs. If the variables are correlated but the distributions are derivable from correlated Gaussian, the pdf of a pair can be found from the joint distribution. The heights of the steps $A_k$ and the values of the nodes $x_k$ are related to the sub-integral of the pdf.           
\subsection{Applicability in finding CDF:} 
It is useful to compare the requirements and complexities of the techniques in relation to inversion of the mgf/chf of a lognormal sum. As noted earlier, the mgf has a wide range of variation in rate and it is necessary to adapt the sampling rate accordingly. Post Widder, Gaver and Gaussian quadrature require summation of weighted mgf near the inverse point as the Equations \eqref{GaverStehfest} and \eqref{equation18}  show. Post Widder uses differentiation of an order $k$ typically greater than twenty at $s=\frac{k}{x}$. The range of $s$ for Gaver and Gaussian quadrature is large for small $x$ and is compressed for large $x$. Gaver technique which uses real arithmetic has a range of the argument of $M(s)$ equal to $N\frac{\ln(2)}{x}$ with a starting point of $N\frac{\ln(2)}{x}$ . Cohen [13] chooses a starting value of $\frac{\ln(2)}{x}$; this makes better use of the low frequency region. The requirement in $s$-space for $x=0.1$ for $N=16$ is as large as $160\ln(2)$. The range for Zakian with $N=10$ is $200$ for $x=0.1$ and $2$ for $x=10$. For both Gaver and Zakiar, the range of $s$ for large $x$ is too large to make effective use of the low frequency end.

\begin{figure}
\begin{center}
 \includegraphics[width=.6\textwidth]{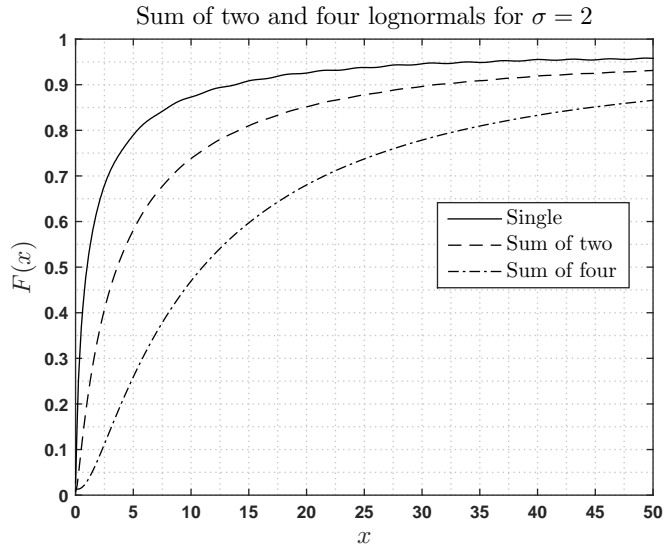}
 \caption{CDFs of sum of lognormals computed by inversion of the chf segmented into four regions.} 
\label{Figure5}
\end{center}
\end{figure}
For similar values of $N$, the moduli of the complex roots of the Gaussian quadrature method [12,13]  have typically a range of $1.3N$ and the $s$-space range has comparable values. For a large value of $x$, the $s$-range is compressed but the requirement of restricting the $s$-values to within the region where $M(s)$ varies fast is hard to meet. The Fourier method uses the entire frequency domain to compute all the values of $x$ ordinarily using uniform grids .This is expensive if one has to accommodate the large changes of rates of variation. Three methods deserve attention. The first is to segment the frequency range in terms of activity noting that the low frequency region has large rate of variation, the far tail has a low rate and the central region has variable activity depending on the variance distribution.  The second is to let the value of the derivative control the segment interval. One can then use exponential integral tables and the simple $\arctan$ expression. These two methods received attention in the present work. 
Figure \ref{Figure5} shows CDFs of a sum of lognormals computed by inversion using four segments. The smallest number of segments needed in Fourier method is two. 
The third is to use multigrid extension suggested by Dalquist [20] where a set of transforms starting at the origin has different values of the highest frequency. One has however to remove repetitions of data. 
\section{Concluding Remarks}
Direct reduced range numerical integration is a convenient tool for finding  the mgf of lognormals on the real line or chf on the imaginary axis and their derivatives. The saddle point method followed by Gauss Hermite quadrature provides insight into the behavior of mgf and derivatives. 

The simplest way to find the inverse of mgf is to use low order Post-Widder/Gaver technique when high accuracy is not demanded. Inversion using Pade approximation is specially useful when almost all components have large variance and the spectrum has dominant low frequency character.  The method employing a sum of complex exponential integrals is simple and attractive because of the availability of tables but special care is needed at low frequencies where the envelope is very close to unity.

Among the three general purpose inversion techniques, viz., Gaver, Gaussian quadrature and Fourier series methods, Gaver and Gaussian quadrature have similar forms and are better equipped to invert if the frequency is not low. Gaver technique is simpler as it uses real arithmetic and is easier to vary the number N of inputs using a recursive algorithm [13]. Both Gaver and Gaussian quadrature techniques  provide pdf and CDF directly. According to the survey by Brian and Martin, the number of terms necessary for Gaussian quadrature is smaller than for Gaver-Stehfest.  The familiar Fourier  series method requires an order of magnitude larger number of terms while providing transforms for all values. However separate computations are needed to find pdf and CDF and it requires modifications for taking care of varying activity rates. The combination of direct numerical integration and Fourier method of inversion appears to be a simple first choice for finding the distribution function. The first two derivatives of the cumulants are directly useful in computing the piecewise sum for CDF and pdf employing tables of exponential integral and complex error function. The simplest approach is to use the arctangent expression based on the first derivatives of the cumulant. One may call attention to the fact that while the distribution at the lower tail is not hard to find, problems remain in finding the distribution at the upper tail. One must find the forward transform at closely spaced small values of frequency near the origin and appropriate inversion methods in this range to meet the accuracy requirements. FW technique, one recalls, has more reliable results at the far tail than at the lower tail.
\section*{Acknowledgment}
The author is grateful to Prof Saswat Chakrabarti for helpful discussions. Thanks are due to Dr. Praful Mankar, Mr. Priyabrata Parida, Mr. Kishore Kumar, and Mr. Dipjyoti Paul for many assistances.

\ifCLASSOPTIONcaptionsoff
  \newpage
\fi


%

%
%
%

\end{document}